\documentclass[10pt,letterpaper,twocolumn]{article} 

\usepackage[draft]{hyperref} 
\usepackage{amsmath}
\usepackage{amssymb}
\usepackage{textcomp}
\usepackage{color}
\usepackage[dvips]{graphicx}

\newcommand{\cre}[2]{\boldsymbol{#1}_{#2}}

\usepackage{cere-ol2_arXive}
\begin{document}
\twocolumn[ 
\title{Narrow-band tunable filter based on velocity-selective optical pumping in an atomic vapor
}

\author{Alessandro~Cer{\`e},$^{1,*}$ Valentina~Parigi,$^{2}$ Marta~Abad,$^{1,3}$ Florian~Wolfgramm,$^{1}$ Ana~Predojevi\'{c},$^{1}$ and Morgan~W.~Mitchell$^{1}$}
\affiliation{$^{1}$ICFO-Institut de Ciencies Fotoniques,\\
Mediterranean Technology Park, 08860 Castelldefels (Barcelona), Spain\\
$^{2}$LENS, Via Nello Carrara 1, 50019 Sesto Fiorentino, Florence, Italy\\
$^{3}$current address: Departament d'Estructura i Constituents de la Mat\`eria,\\Facultat de F\'isica, Universitat de Barcelona, Diagonal 647, 08028 Barcelona, Spain.
}
\email{$^{*}$ Corresponding author: alessandro.cere@icfo.es}

\begin{abstract}
We demonstrate a tunable, narrow-band filter based on
optical-pumping-induced circular dichroism in rubidium vapor.  The
filter achieves a peak transmission of 14.6\%, a linewidth of 80~MHz, and an
out-of-band extinction~$\geq$35~dB.  The transmission peak can be tuned
within the range of the Doppler linewidth of the  D$_{1}$ line of atomic rubidium at
795~nm. While other atomic filters work at frequencies far from absorption,  the presented  technique provides  light resonant with atomic media, useful for atom-photon interaction experiments.
The technique could readily be extended to other alkali atoms.\\
\end{abstract}
]
\maketitle

\begin{figure}
\includegraphics[width=\linewidth]{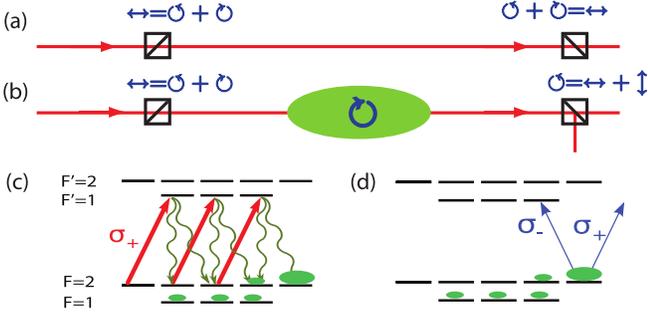}
\caption{\label{img:levels} 
	In (a) and (b) scheme of the polarization interferometer, respectively without and with a dichroic absorber inserted.
	In (c) and (d) the relevant energy levels for  pumping and  probing of $^{87}$Rb together with the induced population distribution.
	Both pump and probe light are tuned on resonance with the 5$^{2}$S$_{1/2}$(F=2)$\rightarrow$5$^{2}$P$_{1/2}$(F'=1) transition.}
\end{figure}

\begin{figure}
	\includegraphics[width=\linewidth]{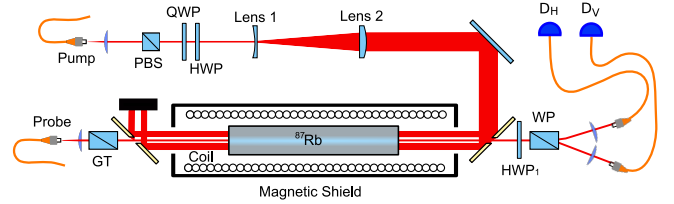}
	\caption{\label{img:setup} Experimental schematics. Two fiber coupled external cavity diodes (not shown) provide probe and pump beams. The probe passes a Glan-Thompson polarizer (GT), the Rb vapor cell, and a Wollaston prism (WP), before being collected  into single mode fibers and detected at D$_H$ and D$_V$. A half wave plate (HWP$_{1}$) is used to balance the interferometer. The pump is magnified to a diameter of approximately 25~mm before entering the cell. Gold coated mirrors with a central 3~mm hole are used to combine the pump and probe beams with minimal polarization distortion.} 
\end{figure}

\begin{figure}
 \includegraphics[width=\linewidth]{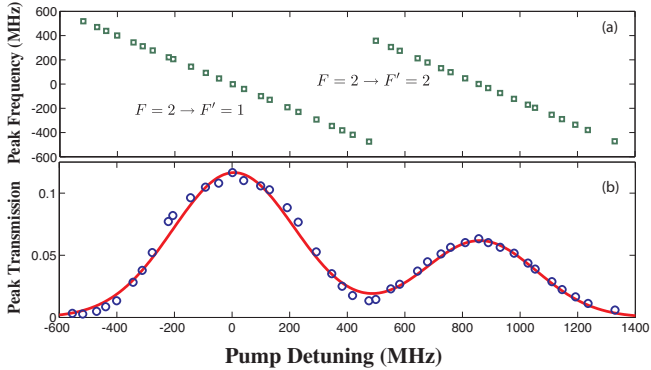}
 \caption{\label{img:tunability}
 Peak center frequency (a) and maximum transmission (b) as function of the detuning of the pump from the  F=2$\rightarrow$F'=1 transition. Every point corresponds to a frame of Media 1. The transmission is well fitted by the sum of two Gaussian distribution (continuous line) corresponding to the Doppler absorption profile of the two transitions indicated. }
\end{figure}

Narrow-band optical filters are a necessary resource in many applications where the  signal to noise ratio is a critical parameter. Applications like free space laser communication~\cite{Junxiong95}, LIDAR~\cite{Fricke-Begemann02,Hoffner05} and generation of narrowband quantum light~\cite{Neergaard-Nielsen07,Bao08,Wolfgramm08} require, together with a high peak transmission, a high out-of-band rejection and frequency stability.
Atomic optical filters based on circular birefringence~\cite{Gayen95,Turner02,He07} have  proven to provide a high  out-of-band rejection and a practicality of use~\cite{Fricke-Begemann02,Hoffner05}.
Other commonly used narrow bandwidth filters are optical cavities. These can  attain high transmission efficiency, narrow bandwidth and  frequency tuning. Because of their comblike transmission spectrum, it is often necessary to use a sequence of two or more cavities with different free spectral ranges~\cite{Neergaard-Nielsen07}.\\
In this letter we present the realization of an atomic filter based on circular dichroism in rubidium.
The velocity selective pumping permits the tuning of the filter over the range of the Doppler width of the atomic transition~\cite{Turner02}.
In contrast to birefringence-based filters, ours uses atomic absorption, thus can be tuned precisely to resonance with the ground-state transitions,
a feature particularly desirable in atomic physics experiments, such as stopped light~\cite{Liu01}.
The design was inspired by  the ``interaction-free measurement'' scheme proposed by  Elitzur and Vaidman~\cite{Elitzur93}.
Their proposal describes a Mach-Zehnder interferometer with the paths balanced so that one output port is dark.  The presence of an opaque object in either path changes the interference and thus increases the probability of a   photon exiting via the dark port.  In our realization, the role of the two paths is played by the two circular polarizations, and the beamsplitters by polarizers as depicted in Fig.~\ref{img:levels}(a) and (b).
The first polarizer selects the initial state in the  horizontal polarization mode ($\cre{\varepsilon}{H} $), corresponding to  an equal superposition of right ($\cre{\varepsilon}{R} $) and left ($\cre{\varepsilon}{L} $) circular polarization modes: $\boldsymbol{E}_{\textrm{in}}=E_{0}(\cre{\varepsilon}{R}+\cre{\varepsilon}{L})/\sqrt{2}$.
If propagating through  a medium with differential absorption for the two opposite circular polarizations and negligible differential dispersion, the state becomes~\cite{Pearman02}: 
\begin{equation}
\cre{E}{\textrm{DC}}=E_{0} \frac{1}{\sqrt{2}}\left\lbrace e^{-\alpha_{R}/2}\cre{\varepsilon}{R}+e^{-\alpha_{L}/2}\cre{\varepsilon}{L}\right\rbrace,
\end{equation}
where $\alpha_{R}(\alpha_{L})$ is real and indicates the absorption for the right (left) circular polarization.
We can rewrite the absorption in terms of differential absorption for the two polarizations defining $\alpha_+=(\alpha_{R}+\alpha_{L})/2$ and $\alpha_-=(\alpha_{R}-\alpha_{L})/2$.
The second polarizer separates two orthogonal linear polarization modes: the originally selected $\cre{\varepsilon}{H}$ and $\cre{\varepsilon}{V}$.
The light intensity exiting at each output of the second polarizer, labeled respectively $H$ and $V$ is then:
\begin{eqnarray}
	\label{eq:prob1}I_H&=& I_{0} e^{-\alpha_+}\cosh^2{\frac{\alpha_-}{2}},\\
	\label{eq:prob2}I_V&=& I_{0} e^{-\alpha_+}\sinh^2{\frac{\alpha_-}{2}}.
\end{eqnarray}
According to Eq.~\eqref{eq:prob2} there is light at  the $V$ output only for $\alpha_-\neq0$.\\
By optical pumping with circularly-polarized light, we generate a circular dichroism feature with a narrow, approximately Lorentzian spectrum.  The pumping scheme is similar to that of Doppler-free polarization spectroscopy~\cite{Pearman02,Harris06}.
We work with the 5$^{2}$S$_{1/2}$(F=2)$\rightarrow$5$^{2}$P$_{1/2}$(F'=1) transition of $^{87}$Rb at a wavelength of 795 nm.
The atomic sample is a hot vapor, so the width of the spectral features is dominated by Doppler broadening.  As shown in Fig.~\ref{img:levels}(c), a $\sigma_+$ polarized pump drives the atomic population toward the F=2,$m_F$=1,2 sublevels.
The intensity is chosen such that the resonant velocity group is efficiently pumped ($I \geq I_{sat}$), but other velocity groups experience little pumping effect.
This creates a sub-Doppler circular dichroism feature~\cite{Happer72,Wieman76}.
The pump can be tuned to address a narrow velocity class of atoms, determining the central frequency of our filter within the Doppler width of the transition.
A similar pumping scheme can be realized for other alkali metals and for different spectroscopic lines, as studied by Harris \textit{et al}.~\cite{Harris06}.\\
A schematic of the experimental setup is shown in Fig.~\ref{img:setup}. The interferometer is composed of a  Glan-Thompson (GT) polarizer  (extinction ratio 10$^{5}:1$) and a Wollaston prism (WP) polarizing beam splitter (extinction ratio 10$^{5}:1$ for both outputs).
A half wave plate (HWP$_{1}$) is inserted between the two polarizers for balancing the interferometer.
The two output ports of the Wollaston prism are coupled into single mode fibers and detected with two silicon photodiodes (D$_{H}$ and D$_{V}$).
The rubidium vapor, a mixture of $^{85}$Rb and $^{87}$Rb (natural abundance),  is contained in a cylindrical glass cell with diameter of 25.4~mm and a length of 15~cm with windows anti-reflection coated on both sides. The cell is heated  to  65\textcelsius, obtaining a pressure of 10$^{-5}$~Torr of rubidium vapor, corresponding to an optical depth of 1.1 on the F=2$\rightarrow $F'=1 transition.
The transmission through  the cell for light far detuned from resonance is 95\%. A bias magnetic field of the order of 100~mG along the cell axis is generated by a coil. A magnetic shielding chamber made of $\mu$-metal reduces the external earth magnetic field to $\leq$1~mG.
The pump light is provided by an external cavity diode laser (ECL).
For optimal background noise rejection, the pump  is directed into the cell counter-propagating with the probe using a large surface gold coated mirror with a 3~mm diameter hole  drilled at its center. 
The pump polarization is set to drive the  $\sigma_{+}$ transitions by a half wave plate (HWP) and a quarter wave plate (QWP).
The angular settings for these wave plates are chosen  taking into account also the polarization rotation introduced by reflection on the mirrors.
The large pump beam waist (approximately 12~mm) ensures a relatively long average crossing time ($>$30~$\mu$s) for the atoms before reaching the center of the cell and interacting with the probe beam.
To achieve velocity-selective pumping neither buffer gas nor paraffin coated cell can be used, thus to efficiently polarize the sample, the pump intensity is much greater than the saturation value of the transition $I_{sat}=1.5$~mW/cm$^{2}$. In this way we obtain the measured values of $\alpha_{R}\approx 5$ and $\alpha_{L}\approx 0.3$.
We use a second ECL as the source of the probe beam. The frequency can be scanned within a range of several GHz around the D$_{1}$ line of $^{87}$Rb.  The probe is collimated into the cell with a diameter of 800~$\mu$m and an intensity $\leq$0.1$~I_{sat}$ is used to  avoid self-pumping effects on the atoms.
\begin{figure}
 \includegraphics[width=\linewidth]{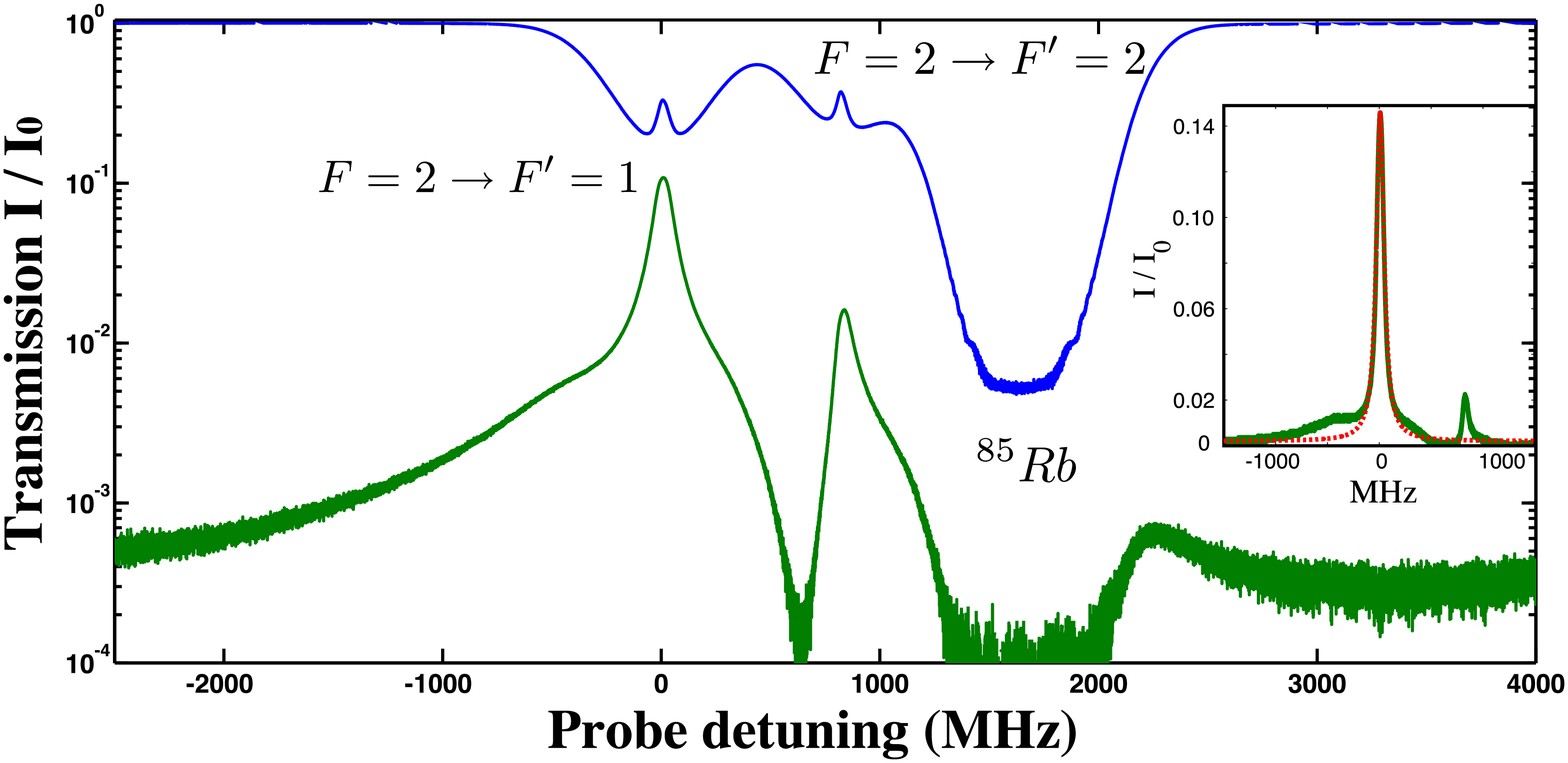}
 \caption{\label{img:average} Experimental transmission spectra of the filter (green). The upper curve (blue) is a reference saturated spectroscopy signal collected at the $H$ output of the interferometer. In the inset a detail of the higher peak and Lorentzian fit (dotted red line).}
\end{figure}
Fig.~\ref{img:average} shows  the spectra collected at the $H$ and $V$ outputs of the interferometer for the pump locked to the F=2$\rightarrow$F'=1 transition. The transmission is corrected for fiber coupling losses. 
The $V$ output presents peaks for the  F=2$\rightarrow$F'=1 and F=2$\rightarrow$F'=2 transitions and a dark background far from resonance. Both peaks are well fitted by Lorentzian distributions in the central part. In the tails of the peaks, the approximation of negligible circular birefringence is not valid anymore. A Gaussian behavior becomes dominant within the Doppler width while for larger detunings the lineshape is Lorentzian again.
From Fig.~\ref{img:average}, the out-of-band extinction ratio is $\geq$~35~dB for probe light detuned by more than 3~GHz, where the effect of the atoms is still appreciable. Further off resonance, the extinction ratio is determined by the quality of the polarizers, their alignment and by the birefringence of the cell windows. In a similar setup, an extinction ratio as high as 70~dB was achieved~\cite{Wieman76}.\\
The maximum observed peak for the F=2$\rightarrow$F'=1 transition is 14.6\% of the input light.
According to Eq.~\eqref{eq:prob2}, a higher optical density would permit a higher transmission but it would also require a higher pump power in order to maintain a good polarization of the whole atomic sample.
Power-broadening by the intense pump beam determines a peak width larger than the 6~MHz natural linewidth of the transition; we measured a FWHM of 80~MHz.\\
The center of the transmission feature can be chosen by tuning the pump frequency within the Doppler profile, as shown in Fig.~\ref{img:tunability}(a)(Media 1). Fig.~\ref{img:tunability}(b) shows the  peak transmission as a function of the pump detuning from the F=2$\rightarrow $F'=1 transition.
The maximum transmission is obtained for the pump tuned to the zero-velocity resonance frequency. For light off resonance, the efficiency decreases together with the reduced optical density of the sample following the Doppler profile of the absorption line. A less efficient pumping is also obtained tuning the pump to the F=2$\rightarrow$F'=2 transition.\\
In conclusion, we have demonstrated a filter for electromagnetic radiation based on an interference technique combined with an appropriate manipulation of the optical properties of an atomic ensemble. We note that the filter could be used with probe modes of various shapes, or with images, provided they are sufficiently collimated. We have measured a peak transmission of 14.6\% and a bandwidth $\leq$~80~MHz, thus attaining a higher transmission and a narrower bandwidth compared to other similar atomic systems~\cite{Gayen95,Turner02,He07}. 
We have also demonstrated frequency tuning  within the Doppler profile width of the absorption line of rubidium.
Moreover, the pumping scheme used  can be extended to other spectroscopic lines of
other alkali atomic vapors, providing a filter for resonant light in atomic optics experiment with hot and cold gases.
This work was supported by the Spanish MEC through Consolider-Ingenio 2010 Project ``QOIT'' and MEC project ``ILUMA'' (ref. FIS2008-01051).
F. W. and A. P. are supported by the Commission for Universities and Research of the Department of Innovation, Universities and Enterprises of the Catalan Government and the European Social Fund.

\end{document}